# Cash and Card Acceptance in Retail Payments: Motivations and Factors

Samuel Vandak, Dr. Geoffrey Goodell

UCL Department of Computer Science

Gower St, London WC1E 6BT

## Abstract

The landscape of payment methods in retail is a complex and evolving area. Vendors are motivated to conduct an appropriate analysis to decide what payment methods to accept out of a vast range of options. Many factors are included in this decision process, some qualitative and some quantitative.

The following research project investigates vendors' acceptance of cards and cash from various viewpoints, all chosen to represent a novel perspective, including the barriers and preferences for each and correlations with external demographic factors.

First, we present the recent history of the growth in card payments, including the digitalisation trend in the UK following the adoption of new regulatory policy in 2012 and the adoption of contactless payments following the SARS-CoV-2 pandemic. Second, we present a breakdown of various card processing fees, which is a major barrier to accepting card payments. We observe that lower interchange fees, limited in this instance by the regulatory framework, play a crucial role in facilitating merchants' acceptance of card payments. The regulatory constraints on interchange fees create a favorable cost structure for merchants, making card payment adoption financially feasible. However, additional factors like technological readiness and consumer preferences might also play a significant role in their decision-making process. We also note that aggregate Merchant Service Providers (MSPs) have positively impacted the payment landscape by offering more competitive fee rates, particularly beneficial for small merchants and entrepreneurs. However, associated risks, such as account freezes or abrupt terminations, pose challenges and often lack transparency. Last, the quantitative analysis of the relationship between demographic variables and acceptance of payment types is presented. This analysis combines the current landscape of payment acceptance in the UK with data from the most recent census from 2021. We show that the unemployment rates shape card and cash acceptance, age affects contactless preference, and work-from-home impacts credit card preference.

## 1. Introduction

The SARS-CoV-2 epidemic has had a profound impact on payment landscape, with a significant increase in card purchases and faster adoption of contactless methods. Limiting the use of physical currency and cash was recommended by governments and national or local health organisations across the globe as a way to slow the virus's spread. The adoption of contactless payment systems gained traction as businesses quickly complied with these recommendations, signalling a paradigm shift in vendor and customer expectations. Health concerns about touching physical currency, as demonstrated by international efforts to disinfect and quarantine banknotes, highlighted this shift. In addition to responding to urgent health needs, the pandemic's acceleration of contactless payments caused a long-lasting change in payment habits for both customers and merchants. The long-term effects of these changes are highlighted by the continued acceptance of contactless payments and the rise in the use of debit cards, even in the wake of the pandemic. Although the pandemic might have hastened the adoption of contactless payments and card payments in general, the rise of such payment methods to the exclusion of cash is not a recent paradigm shift. Over the past ten years, the payment environment in the United Kingdom has experienced a significant shift, with a notable rise in card payments. Important developments in payment technology, such as the launch of the pin-and-chip system in 2004 and the first contactless card in 2007, have been the driving force behind this change. These turning points were essential in improving security and ease of use, setting the stage for the FinTech developments that have shaped modern payment preferences ever since.

## 1.1 SARS-CoV-2 pandemic

During the global pandemic, businesses were urged by governments or specialised health agencies to prefer the acceptance of contactless payment methods while avoiding the physical handling of cash. For example in March 2020, the World Health Organisation suggested that dirty banknotes might be spreading the virus (Gardner, 2020). Just weeks prior to that, the Chinese lenders and vendors were asked to disinfect the notes and safely store them for up to fourteen days before reintroducing them back into circulation (Taylor, 2020). This introduced new barriers to using cash, and thus buyers and sellers alike were forced by the local authorities to use alternative methods. During the first quarter of 2020, the number of payments in China by mobile phone, through third-party mobile payment networks such as Alipay, jumped 187% year over year. Similarly, according to China Internet Network Information Center, the percentage of smartphone owners who use mobile payments rose from 73.5% in June 2019 to 85.3% in March 2020 (China Banking and Insurance News, 2020).

Based on the report by the Bank of England, in 2019, cash payments accounted for only 23% of all transactions, which was a significant drop from approximately 60% a decade ago. However, with the onset of the pandemic in 2020, this figure plummeted further by 35% compared to 2019, with cash only making up 17% of total payments. From 2017 onwards, the usage of cash had already been declining at an average annual rate of 15%, so 2020 marked a rapid acceleration of this trend (Caswell, 2022). According to Buckle (2021), many shops pushed customers to use contactless payments in order to preserve social distancing. To assist with this, the industry upped the spending limit on each contactless card payment from £30 to £45 in April 2020 (Collinson, 2020). A year and a half later, this limit was more than doubled to £100 (Osborne, 2021). Contactless payments remained popular with customers, with more than eight out of 10 individuals now using them. As a result, contactless was used to make more than one in every four payments in the UK in 2020, the first time this threshold was crossed (Buckle, 2021).

An interesting observation is that even after the immediate danger of cash handling during the pandemic faded away, the popularity of contactless payments did not plunge. In November 2020, the Bank of England released a report stating that bank notes pose a low risk of spreading Covid-19 (Osborne, 2020). Despite this, the usage of debit cards rose from 2020 to 2021 by 4.28% from 70% to 73% of all in-store payments (Worldpay, 2022a).

We suggest that the reason for the continued popularity of contactless payments is that by the end of the pandemic, the participants in the market, i.e. vendors and consumers, already built a habit of using this method of payment. Research from Universidade Nova de Lisboa supports the hypothesis that besides the statements by government officials and health bodies, another major reason for people to start using contactless mobile payments was a social influence from the influx of usage in the general population (Zhao and Bacao, 2021). This resulted in a stronger and faster adoption of the technology, which persisted even after the immediate health-related danger had subsided.

### 1.2 Post-2012 trend

It is important to note that the rise of card payments and the decline of cash usage have been ongoing trends that predate the COVID-19 pandemic. While the pandemic has accelerated this transition, specifically in the context of contactless payments, these trends were already well underway before 2020. In fact, the number of card payments made in the UK has increased significantly since 2012 (Payment System Regulator, 2020).

This increase can be attributed to a number of factors, however, the two primary ones include the technological advancements in card payment methods and the regulatory landscape. In 2004, the pin-and-chip method, which made payments more secure than the swipe-and-sign method, was introduced (PaymentSense, 2017). Three years later, Barclays introduced the first-ever contactless card in the UK, making it easier than ever for consumers to make small purchases (Barclays, 2017).

In 2012, the UK government introduced a consumer rights regulation that prohibited businesses from charging customers extra fees for using a particular payment method that exceeds the actual cost incurred by the business for providing that payment option (The Consumer Rights (Payment Surcharges) Regulations, 2012). Before this, specific businesses would charge percentage or fixed extra fees for credit card payments, such as easyJet charging £8 per booking + 2.5%, British Airways charging £4.5 per person, TUI Airlines 2.5% of transaction and Odeon Cinema £0.75 per ticket if the consumer paid with a credit card rather than cash (Shipman, 2011). Following the implementation of the legislation, all payment methods were treated equally from the perspective of the consumer, allowing consumers to use their preferred payment method without incurring any additional costs or barriers, resulting in increased use of digital payment methods such as cards.

## 2. Card barriers - Interchange and Scheme Fees

One of the main barrier to accepting cards are the fees associated with the transactions. The history of charging customers for transactions dates more than sixty years, back when payment card transactions did not use computer technology. Sales clerks authorised and verified the transaction by comparing the card number with a hard-printed book which held a copy of all invalid card numbers. After that, the merchant would send paper drafts to the acquiring bank which then would send the draft to the appropriate network. Following that, the drafts were sorted by issuing bank, packaged, and forwarded to the relevant bank to debit the customer's account (Blakeley and Fagan, 2015).

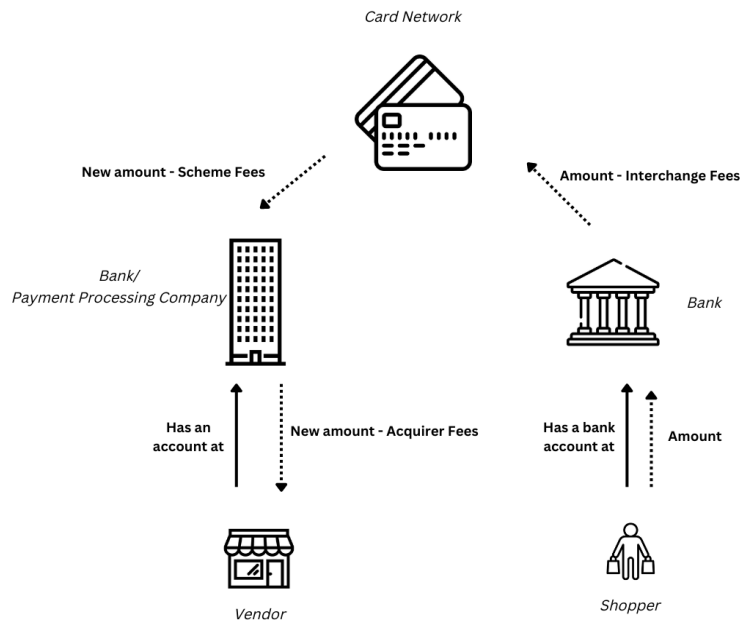

*Figure 1: Fees ecosystem (Vandak, 2023)*

## 2.1 Interchange Fees

Since their introduction in the 1960s, there have been minimal regulations introduced by financial authorities in terms of payment fees. Rather, the fees ecosystem is governed by particular private agreements between the relevant parties - card networks and issuer banks (British Retail Consortium, 2021).

The first recorded case against the payment networks Visa and Mastercard was made by the British Retail Consortium in 1992, after filing an antitrust complaint to the European Commission (British Retail Consortium, 2021). After a generation of pressure from EU businesses and EU business associations and groups, the EU implemented the Interchange Fee Regulation (British Retail Consortium, 2021). Introduced in 2015, it pledged to significantly reduce the cost of payments for merchants (European Commission, 2016). The legislation set a fixed cap for interchange fees in the EU and the UK. Any debit card payment has a set cap of 0.2% interchange fee and any credit card payment has a set cap of 0.3% interchange fee (European Parliament, 2015). The problem with this legislation is that it only limited the cap for consumer debit and consumer credit cards (British Retail Consortium, 2021). Business, corporate, platinum or any other commercial debit and credit cards are excluded from the regulation.

According to the study by EY and Copenhagen Economics the reduction of the interchange fee amounts to one-fifth of the average interchange fee level for EU debit card transactions in 2015. For credit cards, the drop in the interchange fee amounts to around two-fifths of the average interchange charge level for credit card transactions inside the EU in 2015 (EY and Copenhagen Economics, 2020).

As a result, lower interchange fees helped to save merchants in the region a cumulative value of EUR 1,200 million each year, a portion of which will be passed on to customers. Nevertheless, recent merchant reports show that savings are being negated by rises in other expenses, such as scheme fees and commercial card interchange fees. Empirical studies consistently confirm that pass-through onto consumers takes place, but to different degrees for cost increases or decreases and for small or large cost changes. According to the statistical model, consumers receive cost reductions of 66-72% of the merchant's total cost-saving value in the long term (EY and Copenhagen Economics, 2020).

| From/To | Schemes | Issuers | Acquirers | Merchants |
|---|---|---|---|---|
| Schemes | | - 270 | - 280 | |
| Issuers | + 270 | | + 2 680 | |
| Acquirers | + 280 | - 2 680 | | + 1 200 |
| Merchants | | | - 1 200 | |
| **Total** | **+ 550** | **- 2 950** | **+ 1 200** | **+ 1 200** |

*Table 1: Net effect of IF regulation on stakeholders in EUR millions (EY and Copenhagen Economics, 2020)*

This could have been one of the motivations in 2016 for existing vendors who have not yet started accepting card payments to do so or for some vendors to go cashless. As the report points out, since 2015, the acceptance of card-based payments by merchants has dramatically risen, indicated by the number of merchant outlets taking cards and the number of POS terminals. Nevertheless, we find no indication that the rise is greater after 2015 than before 2015, suggesting that the increase may be attributable to factors other than the regulation (EY and Copenhagen Economics, 2020). It is possible that the changes in the fees regulation do not play a significant role in the vendor's thinking process when deciding which payment method to accept; however, lower interchange fee and their pass-through to vendors and subsequently customers might have a positive impact on the vendors who are already accepting cards.

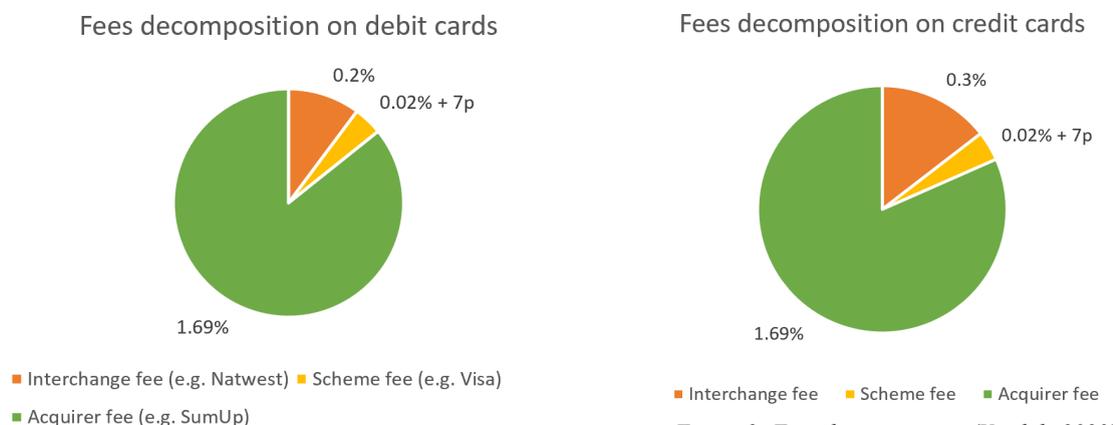

*Figure 2: Fees decomposition (Vandak, 2023)*

The pie charts in Figure 2 show that the interchange fee in fact is a smaller portion of the entire merchant service charge and thus lowering the interchange fee has less of a practical effect on the total cost (Merchant Savvy, 2021). The sources used to create the charts are either from Merchant Savvy or from the average acquiring fees from aggregated merchant service providers SumUp and Square (see Section 2.2, Acquiring/Service Provider Fees), which were obtained on the acquirer's website.

## 2.2 Acquiring/Service Provider Fees

The payment processor or merchant acquirer (acquirer), which is the entity that provides authorization, reporting, and settlement, charges acquirer fees. The acquirer is licensed by card networks and either partners with a payment processor or is a payment processor itself (Optimized Payment Consulting, 2017).

The way that a bank or a financial institution acquires funds for its merchants from a shopper is through a payment gateway service. This service enables retailers to accept payments online, in-app, and in-person via a secure website interface or point-of-sale (POS) terminal system. After the payment getaway verifies the legitimacy of the card, the payment processor sends card information from a merchant's POS system to the card network and banks involved in the transaction (Orem, 2022).

For a merchant to accept a card payment, they have to have an account in a payment processing institution. There are two types of merchant service providers, each with its own distinct characteristics. The first type offers a dedicated merchant account and can act as the merchant's acquirer. The second type offers an aggregated merchant account and, unlike the first type, cannot act as the merchant's acquirer. Further details on each type will be provided in the following paragraphs.

### Dedicated Merchant Service Provider (MSP)

The dedicated merchant service providers commonly act as the vendors' acquiring institutions. They might be closely affiliated with an acquiring bank and offer a custom bank account set up purposely for the vendor's business (Binns, 2023). When the merchant enrols to receive and start using the firm's payment gateway, then the merchant is eligible for a merchant account. With a dedicated merchant account, a merchant can determine custom processing fees rate depending on the size of the business. The rates are specifically based on the monthly volume of sales (Carey, 2021).
According to the consulting group Payments Industry Intelligence, the largest payment processing companies in Europe are WorldPay (from FIS), Barclaycard GPA and JPMorgan Commerce Solutions. More than 80% of all acquired transactions in 2017 were made by these institutions (Payments Card Yearbook, 2018). In this instance, both Barclaycard GPA and JPMorgan Commerce Solutions can utilise their respective UK-domiciled commercial banking license to dedicate a merchant account to their payment-processing customers.

### Aggregated MSP

From the perspective of the consumer, an aggregated MSP serves the role of the main merchant. Smaller merchants act as sub-merchants of the same account and the funds are pooled with the other sub-merchants into the main merchant's account (Paysimple, 2012). The challengers' companies that are popular with small and medium-sized businesses such as Square, SumUp or Zettle are all aggregated MSPs (Lennox, 2020). Due to the aggregated structure, they offer a fairly simple and quick application process, such that the merchants can start accepting the payments on the same day. The transaction rates are fixed, which means that the negotiated fees with dedicated MSP might be lower than fixed fees from aggregated MSP. There exist some serious risks with aggregated MSPs that for marketing reasons are not highlighted on any of the challengers' websites. First, due to a fairly brief, simple and depthless application process, aggregators are more vulnerable to fraud, thus will be more cautious about suspicious behaviour or unusual transactions. As a result, there are more account freezes, holds, and/or abrupt terminations (Fabregas and Bottorff, 2023). Second, aggregators have the option to withhold money at any time. Merchants relinquish financing control and must rely on manually moving funds from the shared merchant account to their custom bank account in a retail bank. This creates a major third-party risk (PaymentOptions, 2020).

The risk associated with aggregated merchant service providers arises from the fact that there are different merchants with the same Merchant Identification Number (MID), meaning their funds are pooled together. As a result, if one merchant in the group engages in fraudulent or illegal activity, the other merchants in the group may suffer as a result (PCI Security Standards Council, 2016). For instance, if one of the group's merchants commits credit card fraud, the acquiring bank may seize the funds, which may result in the freezing or seizure of funds belonging to the group's other merchants during the investigation. Furthermore, if the aggregated merchant service provider is unable to recover the funds from the fraudulent merchant, the other merchants may be forced to bear a portion of the losses (PCI Security Standards Council, 2016).

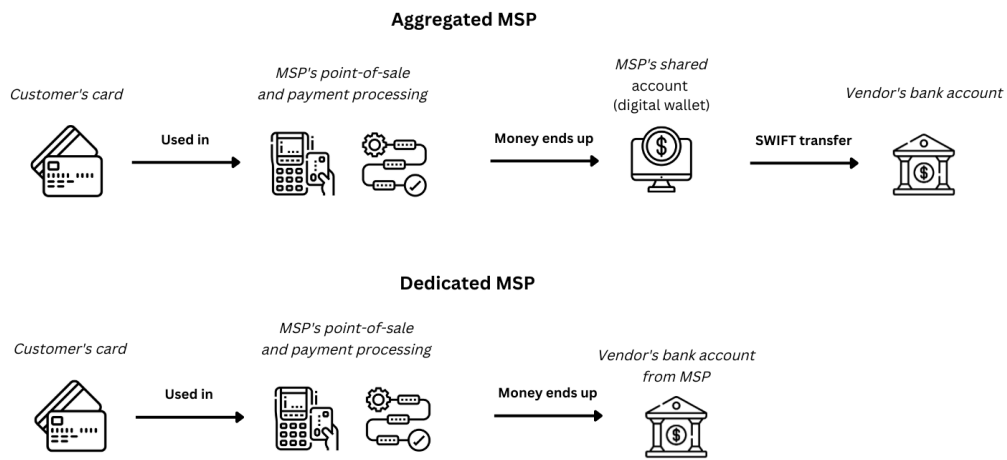

*Figure 3: Types of MSPs (Vandak, 2023)*

*MSPs Limitations*

There are several limiting factors for vendors to consider when choosing between aggregated MSPs ('challenger') and dedicated MSPs ('well-established'). In this section, 'challenger' and 'well-established' are the nouns we are going to use to differentiate between the two as they better illustrate the difference between the two.

| Well-established MSP | Accepted Card Networks |
|---|---|
| Takepayments | Visa, Visa Electron, VPay, MasterCard, Maestro, JCB, AMEX, Diners Club, Discover, UnionPay, GooglePay, ApplePay, SamsungPay |
| Worldpay (FIS) | Visa, Visa Electron, VPay, MasterCard, Maestro, JCB, AMEX, Diners Club, Discover, UnionPay, GooglePay, |

| | AplePay, SamsungPay (WorldPay, n.d.) |
|---|---|

| Challengers MSP | Accepted Card Networks |
|---|---|
| Zettle | Visa, Visa Electron, VPay, MasterCard, Maestro, JCB, AMEX, Diners Club, Discover, UnionPay, GooglePay, ApplePay, SamsungPay (Zettle, n.d.) |
| SumUp | Visa, Visa Electron, Vpay, MasterCard, Maestro, AMEX, Discover, UnionPay, GooglePay, ApplePay (SumUp, n.d.) |
| Square | Visa, Visa Electron, Vpay, MasterCard, Maestro, AMEX, GooglePay, ApplePay (Square, n.d.) |
| MyPOS | Visa, Visa Electron, Vpay, MasterCard, Maestro, AMEX, JCB, UnionPay, GooglePay, ApplePay, SamsungPay (MyPOS, n.d.) |
| Dojo | Visa, MasterCard, Maestro, AMEX, Discover, GooglePay, ApplePay, SamsungPay (Dojo, n.d.) |
| Revolut Reader | Visa, MasterCard, Maestro, GooglePay, ApplePay, SamsungPay (RevolutReader, n.d.) |

*Table 2: Accepted Card Networks.*

As observable, all of the well-established MSPs accept all of the popular card networks. The challenger MSPs all support Visa, Mastercard and American Express, however only one out of six supports Diners Club, two out of six support JCB, three out of six support Discover, and three out of six support UnionPay. The reason for this is that these start-ups would need to dedicate resources to develop connections to these networks. Although the global market share of card brands based on the number of transactions in 2020 was 40%, 32%, 24% and 4% for Visa, UnionPay, Mastercard and others respectively (Statista and The Nilson Report, 2021), in European countries the majority of the transactions are Visa, Mastercard or domestic networks such as Belgium's 'Bancontact'. The market share in these countries is between 90 - 99% (Worldpay, 2022), and in the US the market share is around 85% (Statista, 2022). Since these MSPs mainly operate in these markets, there is simply no need for them to establish connections to the other networks.

| Well-established MSP | Minimum Contract Length (in months) |
|---|---|
| Takepayments | 12 (Darragh, 2022) |
| Worldpay (FIS) | 18 (Bradshaw, 2022) |

| Challengers MSP | Minimum Contract Length (in months) |
|---|---|
| Zettle | 0 (Zettle, n.d.) |
| SumUp | 0 (Lennox, 2020) |
| Square | 0 (Lennox, 2020) |
| MyPOS | 0 (Memotech, n.d.) |
| Dojo | 6 (Sorensen, 2022) |
| Revolut Reader | 0 (RevolutReader, n.d.) |

*Table 3: Contract length*

While the majority of the challenger MSPs do not require a minimum contract at all, the well-established dedicated MSPs usually have between 12-24 months of minimum contract depending on the product. In this analysis, we have only considered the products with the shortest minimum contracts, yet none is below 12 months. This might be the primary reason why many SMEs or starting vendors prefer using services by SumUp, Square, Zettle and others. For these companies, small vendors are the target customers. As Enlyft, an account intelligence firm, notes, Square is most often used by companies with fewer than 10 employees and yearly revenue under 5m USD (Enlyft, n.d.). One in five small businesses fail within the first year in the UK, and the figure rises to 60% when it comes to the first three years (Horne, 2022). New retailers often face challenges in their early years due to the high risk of new venture failure. This phenomenon is known as Lindy's effect, which states that the longer something has been around, the longer its future life expectancy is likely to be. In other words, the longer a business survives, the more likely it is to continue to survive (Eliazar, 2017). In mathematics, the theorized phenomenon follows the Pareto probability distribution (Eliazar, 2017). In the UK, 60% of small businesses fail within their first three years (Horne, 2022). New enterprises are specifically vulnerable to an early failure, and therefore early-stage vendors can't risk signing up for a lengthy contract.

*MSPs Fees Analysis*

The various fees can be categorized in three different ways when it comes to card-present transactions through point-of-sale. First, there are transactions in which local domestic cards (UK cards) were used, second, there exist transactions with non-domestic foreign cards being used and third, transactions where American Express cards were used. In all of these three categories, the fees change distinctively.

Some of the MSPs have also a minimum flat fee, usually between 3-10p per transaction. To further understand the impact of this seemingly low flat pence fee, there are two analyses for each category. The first one is a small transaction worth £2, the second one is a larger transaction worth £100.

Each MSP has a horizontal line that illustrates the income level. An empty circle shows the value of the theoretical real income - the original value of purchase. A full circle or the area between two full circles show the range between the levels of real income after the MSP fees are deducted. The full breakdown of the fees, in percentage figures, can be found in the appendix (see Appendix A).

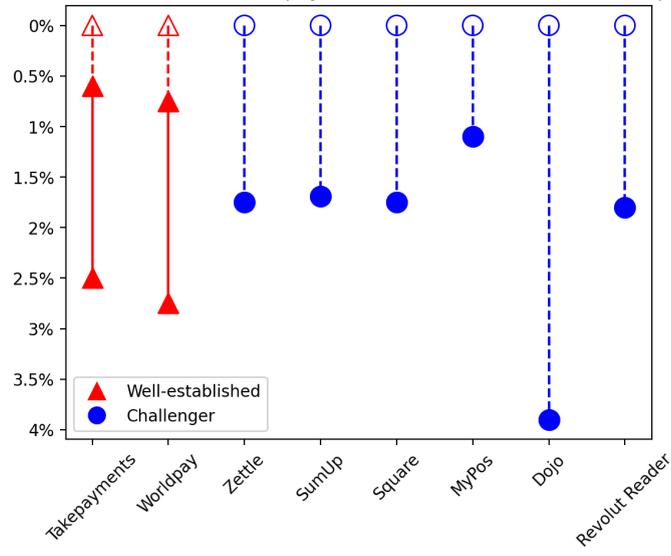

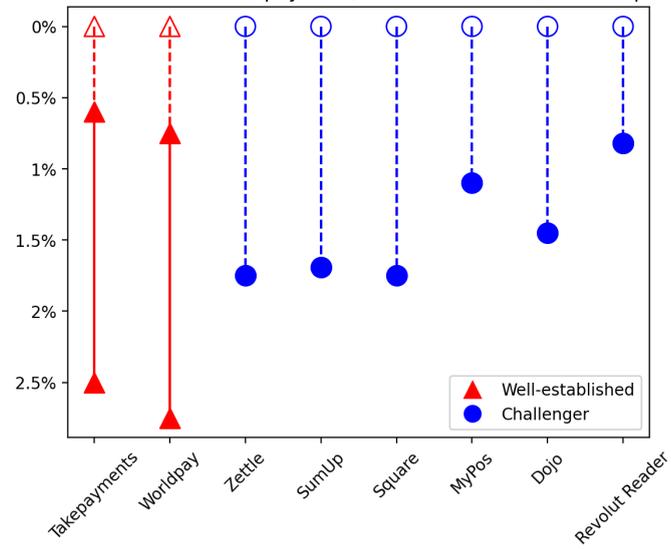

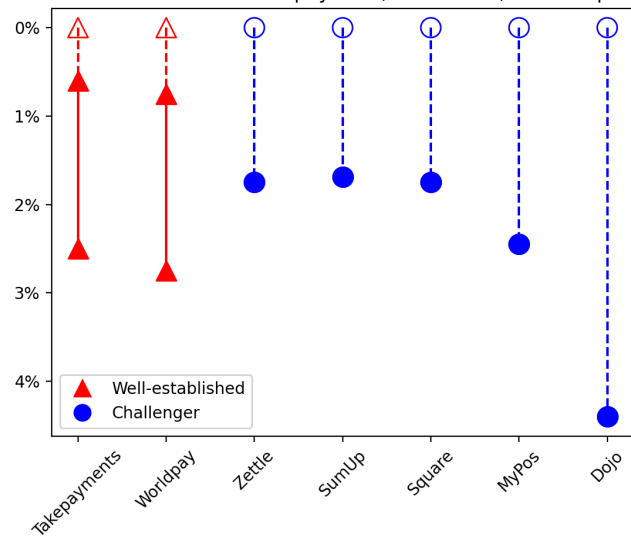

There are several trends from the obtained data. Firstly the main difference between the well-established dedicated MSPs and the challenger aggregated MSPs is that while the first has a bespoke fees rate, the second has a transparent flat ad valorem rate. The bespoke rate might depend on multiple factors that are not publicly available and considered part of the company's confidential process. However, the most important factor is the business' average transaction volume and value (Darragh, 2023). Given these criteria, retail establishments such as cafes and convenience shops are more likely to have higher fees than businesses with higher minimum transaction requirements, such as restaurants.

The second trend is that there are two entities which besides having a fixed ad valorem transaction fee also operate on a flat pence rate. This means that the total fee will be a percentage of the transaction plus a fixed amount in pence. If the value of the transaction is low and the vendor operates in low-value item sales such as a coffee shop or an independent grocery store, the vendor will lose a substantial amount of the purchase value. Therefore these vendors are more likely to avoid Dojo and Revolut Reader. Both of these have however lower ad valorem fee rates and therefore when it comes to higher value items, where the absolute extra pence rate is negligible, the real income is higher than if the vendor used Square, SumUp and Zettle.

American Express, compared to other card issuers, generates a larger portion of its revenue from fee income due to their unique business model. Unlike others, it relies on transaction fees charged to merchants and annual fees charged to its customers rather than interest income (Reiff, 2021). For instance, the flagship Platinum card requires cardholders to pay their entire balance each month but charges a $700 annual fee for the card's benefits (American Express, 2023). Additionally, it incurs higher expenses compared to other networks due to the many benefits that it offers to its cardholders, such as exclusive access to airport lounges, concierge services, and rewards programs (Parker, 2023).

When it comes to accepting American Express cards, the situation varies from provider to provider. Some providers like Zettle, SumUp, and Square have fixed rates for all the card networks, which means there is no preference for vendors if the customer uses Visa, Mastercard or Amex. On the other hand, MyPos and Dojo have higher rates for Amex transactions.

Another special case is Revolut Reader. Revolut Reader is a newly established product by Revolut, launched in June 2022 (Revolut, 2022). Although being labelled as a challenger MSP, instead of being an aggregate MSP, Revolut Reader is a dedicated MSP as it automatically sends the payment to the vendor's Revolut bank account. Due to a recent launch, Revolut Reader has an immense competitive advantage due to very low transaction fees. However, the fees might change once Revolut acquires more merchants and a critical mass of customers.

## 3. Quantitative Study

The United Kingdom is one of the most divided countries when it comes to the payment landscape. On one hand, we can observe a very swift and apparent transition towards a cashless society. In 2021, more than 23 million people did not use cash at all as a form of payment (Jones, 2022). During the same time, the percentage share of payments made by debit and credit cards in the UK was 8% higher than in EU peers (50.8% vs 47%) (UK Finance, 2020) (European Central Bank, 2021). Overall, the usage of cash is at its lowest point accounting for only 17% of all transactions (UK Finance, 2021).

Researchers predict that within the next ten years, the usage of cash will drop further, such that notes and coins would only be used in 6% of all transactions in the UK (Jones, 2022). This shows that the adoption of cashless payment methods in the UK is high and is still steadily growing.

On the other hand, according to the Financial Conduct Authority, there are 1.3 million adults in the UK that do not have a bank account (FCA, 2018). There are multiple reasons why individuals are not account holders, but they can be categorized mainly into two categories, voluntary and involuntary. A third of 'unbanked' persons voluntarily choose to not have bank accounts; for example, they previously had a bank account but no longer want one (FCA, 2018). The rest are involuntarily unbanked for reasons such as incompetence to open a bank account, needing help to provide proof of identity, or inability to provide proof of residency. The overwhelming majority of people without a bank account, about 90%, are from low-income households (Green, 2022). Furthermore, approximately eight million adults in the UK rely on cash and use cash as a means of payment at least once a day (Green, 2022). Despite the UK being on a path to the cashless payment landscape, some of the demographic groups, such as low-income households, are not ready for this transition.

As discovered during the qualitative interviews, one of the reseasons why vendors accept and prefer a specific payment method is due to customers' convenience and preference. In this case, it was mostly contactless payment. There is a relationship between vendors' preferences and customers' preferences. This relationship thus helps to dictate what payment methods the vendors accept. As mentioned in the previous paragraph, there exist specific demographic groups that do not possess a bank account and only pay with cash. In this part of the research, we will investigate other relationships between various demographic variables in the areas where the vendors are located and vendors' acceptance of payment methods.

### 3.1 Self-collected data

Self-collected data refers to the vendors' preferences data that was collected manually throughout the research. The method of data collection was a simple questionnaire form. The order of questions follows a specific pattern:

1. Defining questions - Postcode and Location of the business
2. Preference questions (questions regarding the vendors' preferences)
   a. What payment methods do you accept?
   b. What payment method is the most used one?
   c. What payment method do you prefer to accept the most?

The sectors targeted were the hospitality and retail sectors. These sectors will offer a good representation of local customers. Since our census-collected data comes from the geographical location of the business, it is important to analyse businesses whose customers are likely to be local, therefore the majority of the data is from coffee-related businesses, pubs, bars and small independent specialised and convenience stores.

There were several methods used to share the form and collect the data. The first method was through email outreach: Following a manual look-up for various businesses on Google Maps, a data sheet with thousands of email entries was formed. Additionally, businesses were contacted through their public phone numbers and asked questions during a call, as well as visited in person and asked to participate in the study.

The sampling method used to obtain email addresses and telephone contact details of the retailers involved selecting locations on Google Maps and searching for specific sub-sectors such as coffee shops, pubs, and restaurants. Following the generation of search results, the websites of the retailers were visited to acquire the contact information. While this sampling method has limitations, it is a simple and cost-effective way of gathering information on a large number of retailers.

On the other hand, due to the researcher's location, the in-person outreach had to be limited to a single city: London. To diversify the responses, three distinct London boroughs were visited: Westminster, the City of London, and Camden. The information gathered was evenly distributed across these three areas, each with its own set of demographics and characteristics.

Table 4 shows the number of collected forms by each method.

| Method | E-mail Outreach | Telephone Outreach | In-person Outreach |
|---|---|---|---|
| Contacted/Approached | +- 1,000 | 105 | 40 |
| Collected samples | 150 | 25 | 12 |

Table 4: Descriptive statistics of data collected

### 3.2 Census-collected data

In the UK, the census is held every ten years and provides a snapshot of all the people and households in England and Wales. The most recent Census Day was on Sunday, March 21, 2021, with the first results released on 28th June 2022 (Office for National Statistics, 2022). The data that was used in this research was published between June 2022 and January 2023

The following data sources were used for the analysis:

1. Demography and Migration - TS007 - Age by single year of age
2. Work and Travel - TS066 - Economic activity status (unemployment rate)
3. Work and Travel - TS061 - Method of travelling to work (work from home rate)
4. Population density using the data from TS007 and geographical boundaries data

### 3.3 Data Cleaning

The biggest data cleaning challenge was related to how the census is conducted in the UK. The data from the census is released in multiple geographical forms. On the top of the hierarchy is the data by country (England, England + Wales, Wales), on the other hand, the smallest geographical unit is output areas. We will use the Middle Layer Super Output Areas (MSOA) in this project. MSOAs have a minimum of 5000 residents and 2000 households with an average population size of 7800. They fit within local authority boundaries (Staffordshire County Council, 2011). Using this layer will allow us to have a larger sample in order to obtain a better representation of what area the business is located in such as urban/rural, high-income level/low-income level, etc.

To match the census data with the data collected from retailers, the postcodes (key field) were converted to MSOAs using a file that contains all of the UK's 1.8 million postcodes and their respective area codes including the MOAS (Office for National Statistics Geography, 2022).

*3.4 Implementation*

While the census data contains continuous independent variables, the vendors' data are categorical variables. For instance, an answer to a question asking 'What is the most used payment method' the answer could be 'Debit card', 'Cash', and so on. In order to create a functioning model, the categorical variables were transformed into binary variables using modified dummy variable encoding. Modified encoding means that in this instance a special type of grouping will be used or that some encodings might be omitted due to the lack of data or data bias. These instances are explained fully later.

The encoded data can be categorized accordingly:

- Acceptance
    - Does accept cards?
    - Does accept cash?

- (Highest) Usage
    - Is the most used method a contactless mobile payment (digital wallet)?
    - Is the most used method a debit card?
    - Is the most used method a credit card?
    -
- Preference
    - Does the vendor prefer a debit card?
    - Does the vendor prefer a credit card?
    - Does the vendor prefer cash?

The modified encoding constituted several changes in the categorization example. First, instead of asking whether the vendor supports debit cards and credit cards independently, these two questions were merged, and the input for the model is whether the vendor accepts cards and whether the vendor accepts cash, i.e. whether the retailer is cardless or cashless. While retailers can choose to accept debit cards and not accept credit cards, the data showed that all the retailers that accept at least one of them, in reality accept both. Another modification was that the 'Is the most used method cash' data was omitted. This is because none of the vendors that accept both cards and cash had claimed that cash is more used. Cash was predominantly the most used payment method for only cardless retailers. Finally, in the preference section, the encoding for whether the vendor prefers contactless payments were not used. This is because the contactless payment in its essence, from the vendor's point of view, is like any other debit payment since we have shown in the previous section that all the MSPs (Merchant Service Providers) accept Apple Pay and Google Pay. This encoding was only used in the usage section as that one is customer oriented where the preference might matter for instance for a specific demographic age groups.

After encoding the data, the model was created by using logistic regression. Logistic regression is a technique wherein both continuous and categorical data is encoded to binary data using dummy variable encoding. This is because logistic regression is a supervised learning algorithm that can be used to predict binary outcomes, such as whether an event will occur or not. The logistic function is used to model the probability of the outcome of interest based on the input features, which can be both continuous and categorical (Albon, 2018, pp.308–318).

One of the main advantages of logistic regression is that it is easy to interpret and understand the relationship between the input features and the outcome. The coefficients of the logistic regression model can be used to quantify the strength and direction of the relationship between each input feature and the outcome (V Kishore Ayyadevara, 2018).

### 3.5 Bias and limitation

First of all, the low response rate and having approximately 200 data inputs raises concerns about the sample's representativeness, since those that answered might not fairly represent the general population of vendors. However, this is due to limitations in capacity, outreaching options, and funding to undertake a comprehensive large-scale study on the same level as professional agencies.

Second, vendors who do not have access to or choose not to use the internet and email would not be able to be reached as email was the primary mode of data collection. Furthermore, the information gathered could favour people who are more familiar with technology and at ease doing online surveys.

To reduce this bias, some of the data were gathered by in-person outreach and calling the business numbers. However, this has a bias on its own as only the vendors in specific locations were asked to participate.

### 4. Quantitative Study Results

In order to determine the significant predictors of the outcome variable and their effect in our logistic regression model we use the odds ratio. The odds ratio is a useful statistical tool in logistic regression analysis because it indicates the strength and direction of the association between the predictor variables and the outcome variable (Hilbe, 2015). It enables us to estimate the change in the probability of the outcome variable for a unit increase in the predictor variable while controlling for the effects of the model's other variables. For instance, if the odds ratio for a predictor variable is 2, it signifies that a one-unit increase in that variable results in a doubling of the probabilities of the outcome variable. When dealing with binary outcomes, the odds ratio is especially useful because it provides a clear and interpretable measure of the effect size (Hilbe, 2015).

The charts also contain p-values that are a valuable tool in hypothesis testing using the null hypothesis. P-values provide a measure of the strength of evidence against the null hypothesis, allowing us to determine whether to reject or fail to reject it. By including p-values in charts, we can visually represent the significance of our results and make informed decisions based on statistical evidence. We use a 5% significance level meaning we reject the null hypothesis if the p-value is less or equal to 0.05.

*4.1 Relationship with average age data*

*Motivation:*
There are vast differences in how often different age groups use cards and cash. According to the global YouGov survey, 47% of young adults between the age of 18 and 24 have made a cash payment in the last three months. This is significantly less compared to 70% of people aged 55 and older. In contrast, while 29% of late generation Z made a contactless mobile payment in-store using a digital wallet, only 18% of older adults (55+) did so (YouGov PLC., 2020). Thus in areas with a younger population, retailers might have a higher percentage of people using contactless as their most used method. However, in an area with an older population, credit card and cash usage would be higher than contactless debit.

*Acceptance $H_0$* : The average age in the given area has no effect on the vendor's acceptance of card payments or acceptance of cash.

*Usage $H_0$* : The average age in the given area has no effect on whether the most used payment method is contactless, debit card or credit card.

*Preference $H_0$* : The average age in the given area has no effect on whether the vendor prefers debit card payment, credit card payment or cash payment.

***Models:***

<u>Acceptance</u>

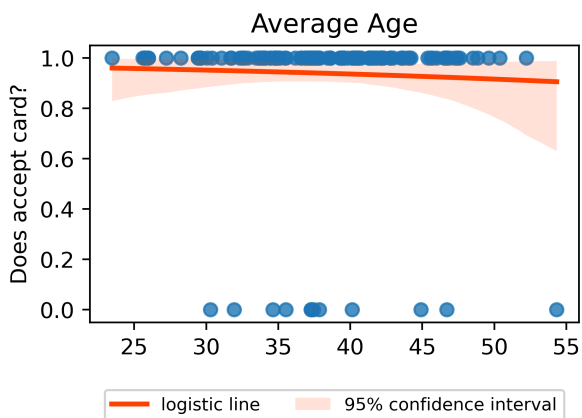

Odds Ratio (OR): 0.970681
p-value: 0.5914

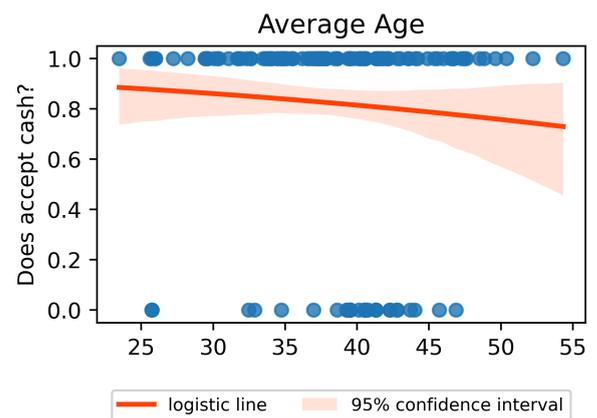

Odds Ratio (OR): 0.966551
p-value: 0.3298

<u>Usage</u>

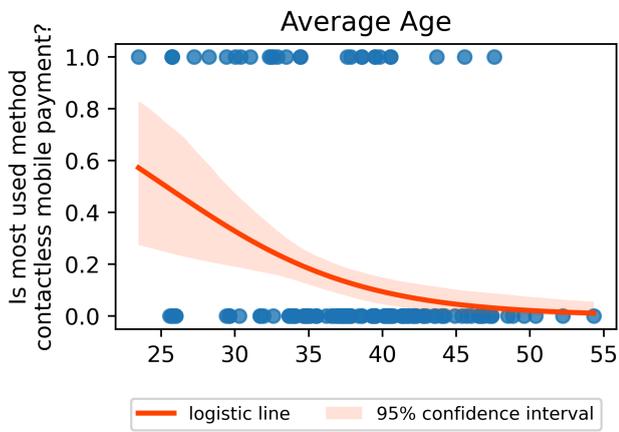

Odds Ratio (OR): 0.857015
p-value: <0.001

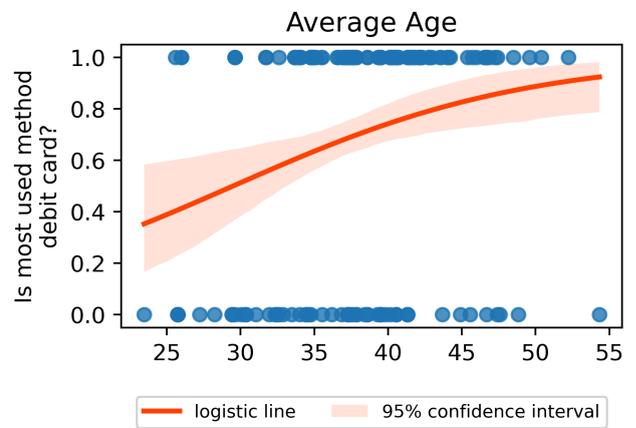

Odds Ratio (OR): 1.105817
p-value: 0.0011

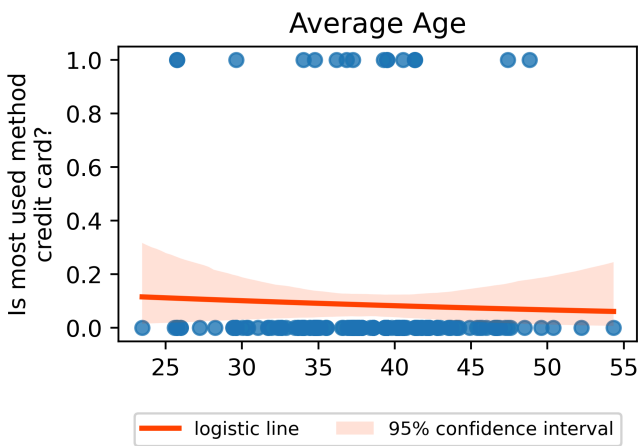

Odds Ratio (OR): 0.9778
p-value: 0.6276

<u>Preference</u>

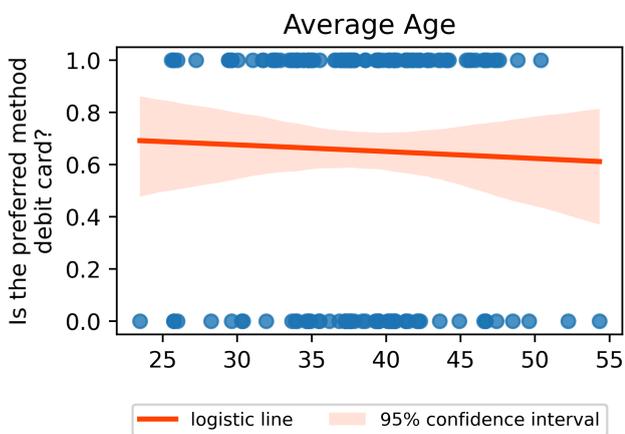

Odds Ratio (OR): 0.988606
p-value: 0.6788

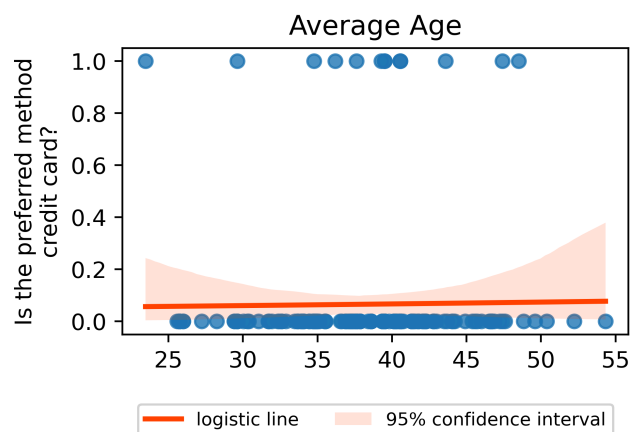

Odds Ratio (OR): 1.01078
p-value: 0.8396

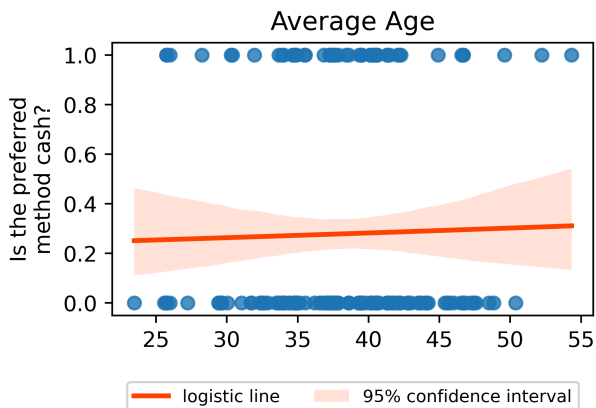

Odds Ratio (OR): 1.009631
p-value: 0.7438

*Acceptance $H_0$*: We fail to reject the null hypothesis.

*Usage $H_0$* : We reject the null hypothesis.

*Preference $H_0$* : We fail to reject the null hypothesis.

The data reveals that the likelihood of the most often used payment option becoming contactless decreases as the average age in the area of the business rises. P-values at levels of around 0.001 are used to support this association, indicating its statistical significance. The odds ratio of 0.85 means that for every one-year increase in the average age, the odds of the most used payment method being contactless decrease by 15%. This conclusion may be explained by the fact that older consumers are less inclined than younger consumers to utilise or trust new technology and rather prefer to pay with debit cards or cash. In regards to debit cards, the relationship is reversed which supports this hypothesis. In terms of preference and acceptance of payment methods by vendors, there is no statistical significance with any of the factors presented.

## 4.2 Relationship with unemployment level data

*Motivation:*

Unemployment is one of the barriers to paying with a card. According to the Federal Reserve Bank of Boston, unstable employment changes consumers' ability to acquire and pay with a credit card (Cole, 2016). Retailers in areas with high unemployment levels could be therefore more likely to be cash-only as their local customers do not have the need and possibility of card payment.

*Acceptance $H_0$* : Unemployment in the given area has no effect on the vendor's acceptance of card payments or acceptance of cash.

*Usage $H_0$* : Unemployment in the given area has no effect on whether the most used payment method is contactless, debit card or credit card.

*Preference H₀* : Unemployment in the given area has no effect on whether the vendor prefers debit card payment, credit card payment or cash payment.

**Models:**

<u>Acceptance</u>

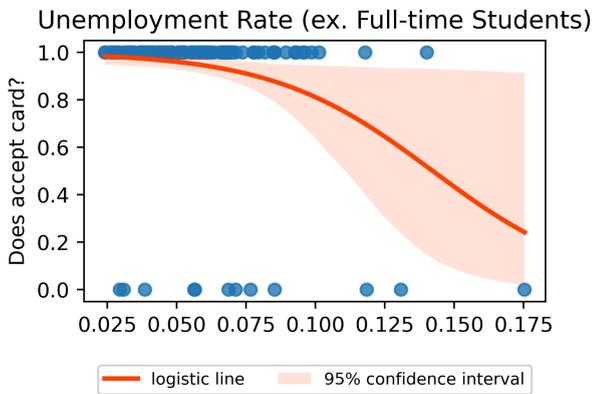

Odds Ratio (OR) 0.761725
p-value: 0.0013

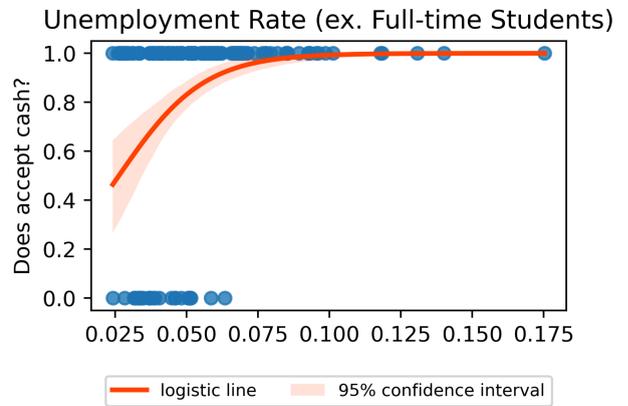

Odds Ratio (OR) 1.535894
p-value: <0.001

<u>Usage</u>

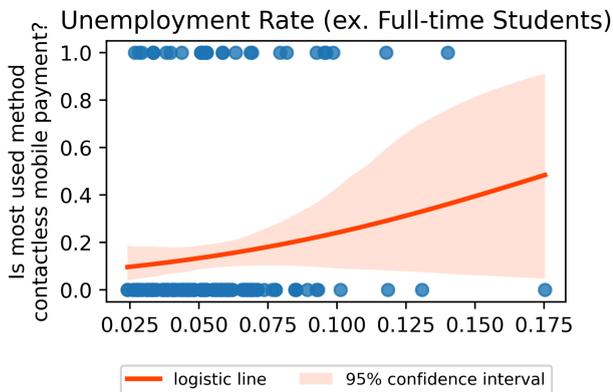

Odds Ratio (OR): 1.198834
p-value: 0.0862

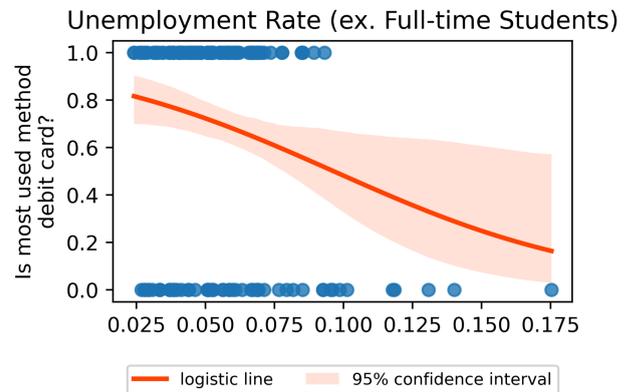

Odds Ratio (OR): 0.686381
p-value: 0.0075

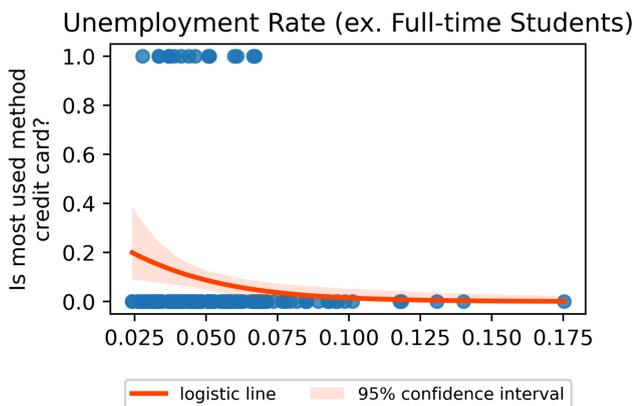

Odds Ratio (OR): 0.855744
p-value: 0.0473

<u>Preference</u>

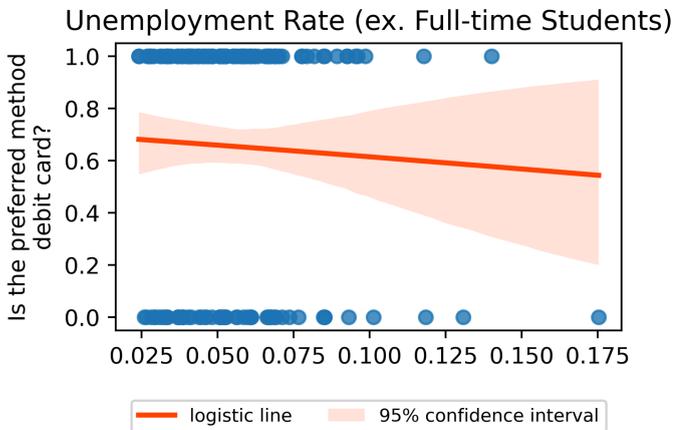

Odds Ratio (OR): 0.928079
p-value: 0.5847

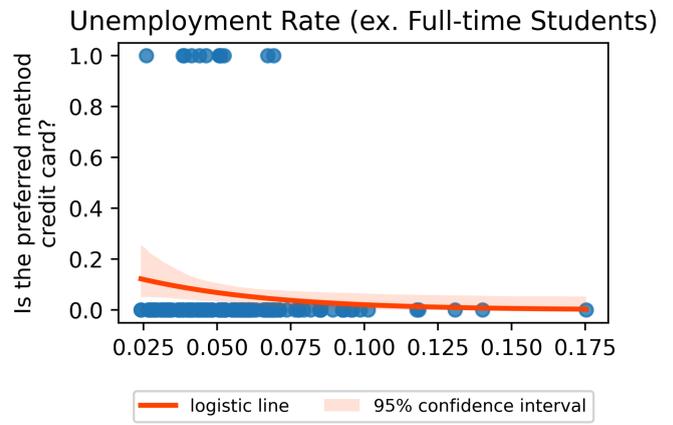

Odds Ratio (OR): 0.912564
p-value: 0.1955

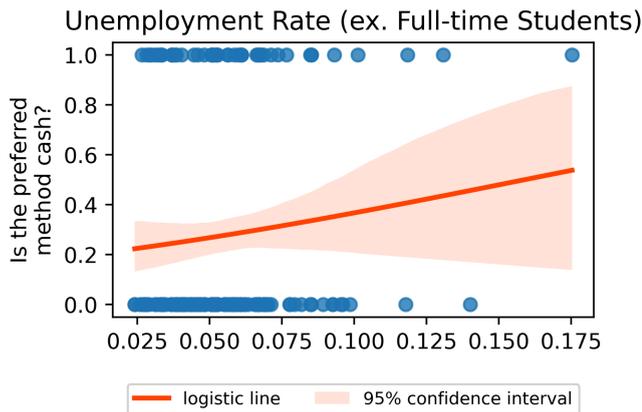

Odds Ratio (OR): 1.179665
p-value: 0.206

*Acceptance $H_0$*: We reject the null hypothesis.

*Usage $H_0$*: We reject the null hypothesis.

*Preference $H_0$*: We fail to reject the null hypothesis.

The higher the unemployment rate in the area, the more likely the vendor accepts cash. This can be explained by the fact that people who are underbanked or unbanked frequently prefer using cash as a form of payment as they do not have access to credit or debit cards.

In terms of customers' usage, unemployment is likely to affect the usage of debit and credit cards. The odds ratio of 0.69 and 0.86 for debit and credit cards respectively means that for every unit increase in the unemployment rate in the area, the odds of debit card and credit card being the most used payment method decrease by a factor of 0.69 (or 31%) and 0.86 (or 14%).

*4.3 Relationship with commute data*

*Motivation:*

The rate of commuters who use public transport depicts the urban landscape of the geographical location. No ownership of a car in urban conurbation is three times higher than in rural towns and fringes. In these conurbations, the commute to work is largely dependent on public transport services (Department for Transport, 2021). There might be differences in the retail vendors accepting and preferring different methods depending on whether they reside in rural or urban areas and thus this data was used.

*Acceptance $H_0$*: The commute by public transport rate in the given area has no effect on the vendor's acceptance of card payments or acceptance of cash.

*Usage $H_0$*: The commute by public transport rate in the given area has no effect on whether the most used payment method is contactless, debit card or credit card.

*Preference $H_0$*: The commute by public transport rate in the given area has no effect on whether the vendor prefers debit card payment, credit card payment or cash payment.

**Models:**

<u>Acceptance</u>

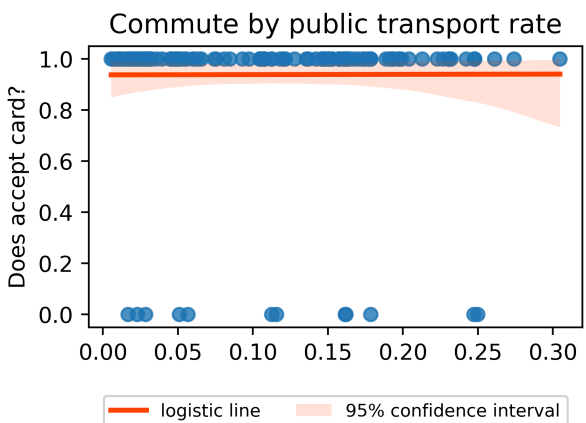

Odds Ratio (OR): 1.008265
p-value: 0.9693

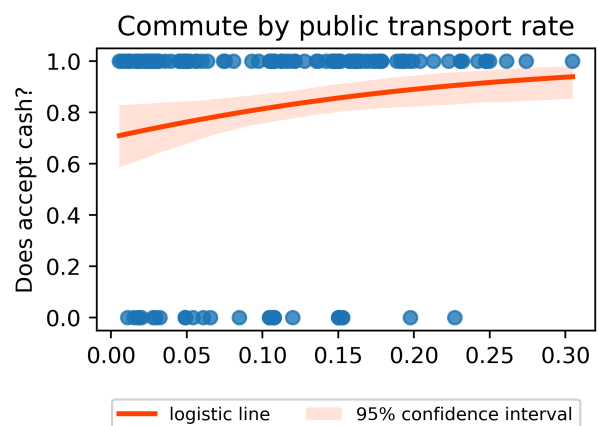

Odds Ratio (OR): 1.305748
p-value: 0.4044

## Usage

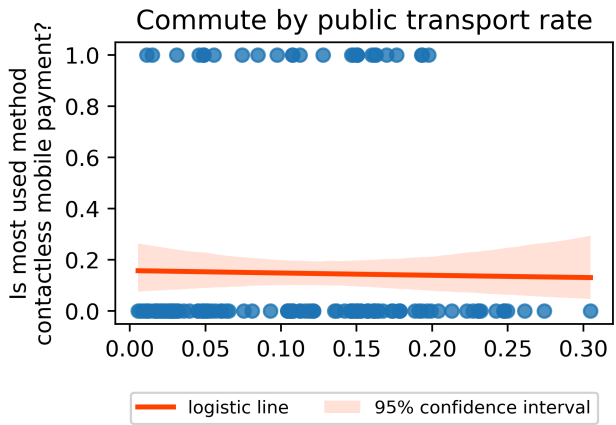

Odds Ratio (OR): 1.305748
p-value: 0.4044

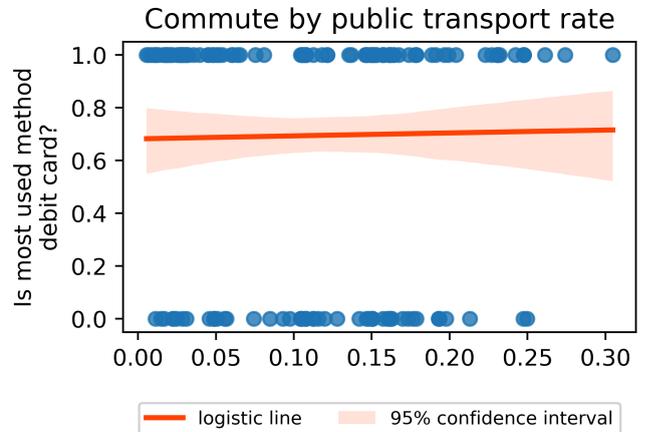

Odds Ratio (OR): 1.087261
p-value: 0.819

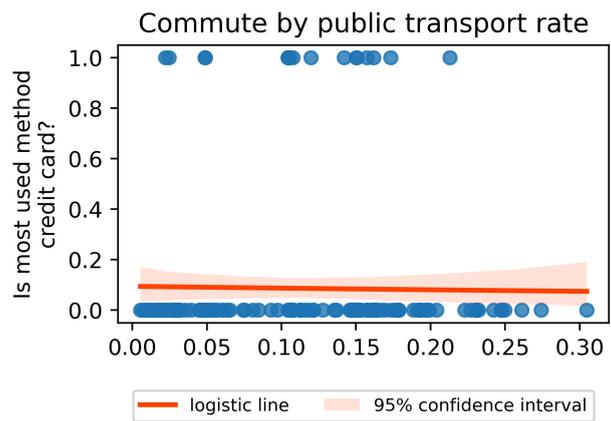

Odds Ratio (OR): 0.945534
p-value: 0.8211

## Preference

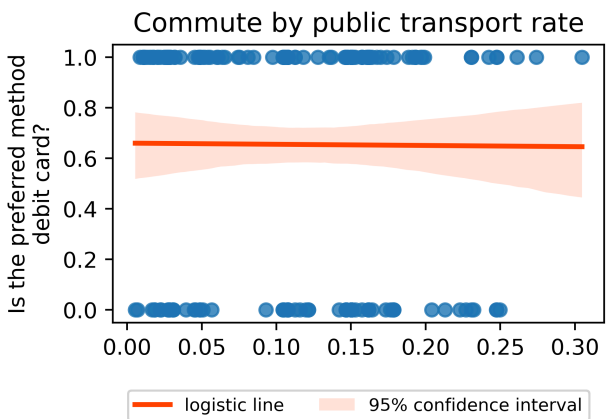

Odds Ratio (OR): 1.305748
p-value: 0.4044

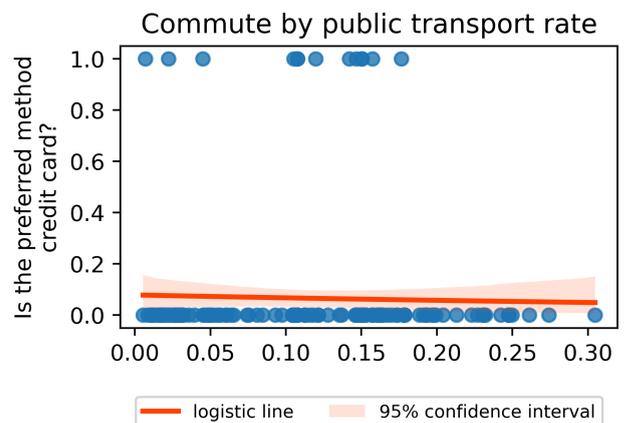

Odds Ratio (OR): 0.917423
p-value: 0.6983

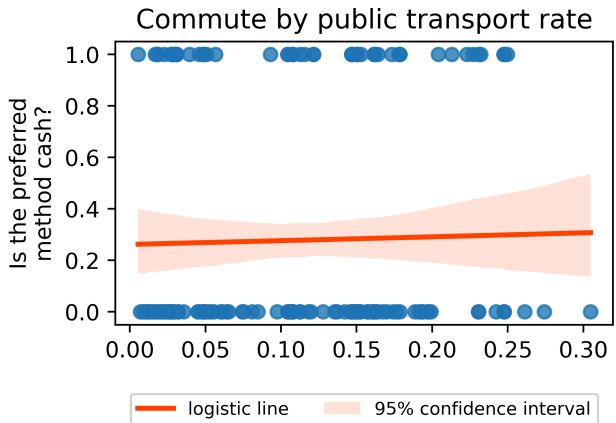

Odds Ratio (OR): 1.118699
p-value: 0.7549

*Acceptance $H_0$*: We fail to reject the null hypothesis.

*Usage $H_0$* : We fail to reject the null hypothesis.

*Preference $H_0$* : We fail to reject the null hypothesis.

There is no statistical significance regarding the effect of public transportation usage and the retailers' acceptance and preference of cash and card methods and the usage of specific payment methods by customers.

### 4.4 Relationship with work-from-home data

*Motivation:*
According to the studies, the work-from-home rate is significantly ($p < 0.01$) correlated with income and annual salary. This effect was even more significant during the Covid pandemic (Nwosu, Kollamparambil and Oyenubi, 2022). Since the census took place in 2020/2021, the effect of the pandemic is included. Demographics with white-collar jobs that were most likely remote might have a different preferred type of payment method and the retailers in the given areas are more likely to represent that too.

*Acceptance $H_0$*: The work-from-home rate in the given area has no effect on the vendor's acceptance of card payments or acceptance of cash.

*Usage $H_0$* : The work-from-home rate in the given area has no effect on whether the most used payment method is contactless, debit card or credit card.

*Preference $H_0$* : The work-from-home rate in the given area has no effect on whether the vendor prefers debit card payment, credit card payment or cash payment.

*Models:*

<u>Acceptance</u>

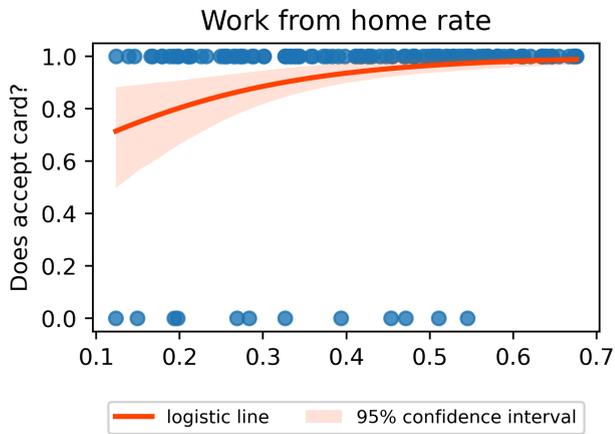

Odds Ratio (OR): 4.101858
p-value: 0.0014

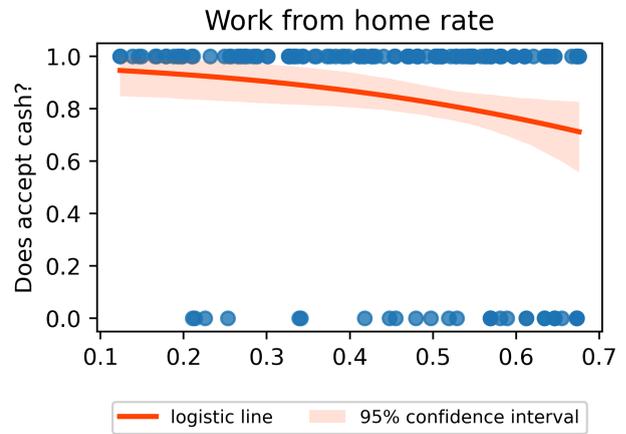

Odds Ratio (OR): 0.298925
p-value: 0.0182

<u>Usage</u>

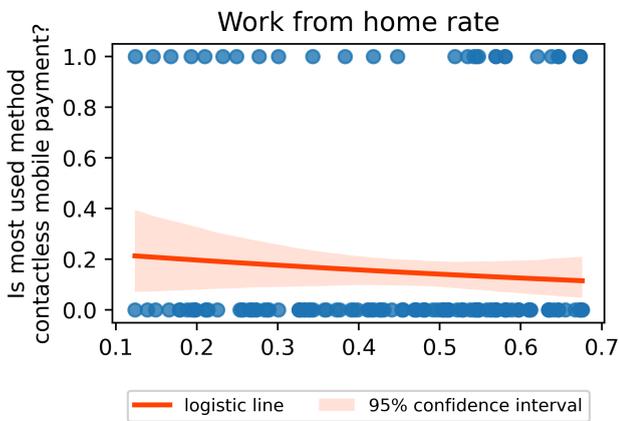

Odds Ratio (OR): 0.611596
p-value: 0.304

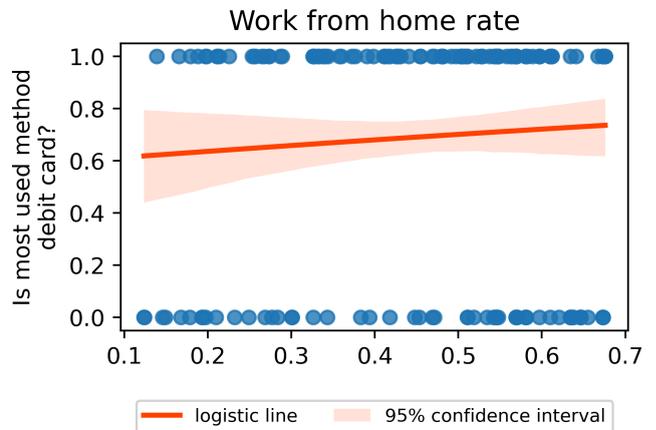

Odds Ratio (OR): 1.617476
p-value: 0.3349

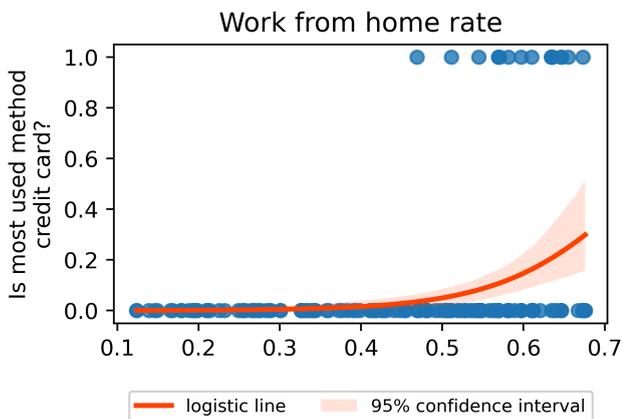

Odds Ratio (OR): 4.875879
p-value: 0.0016

<u>Preference</u>

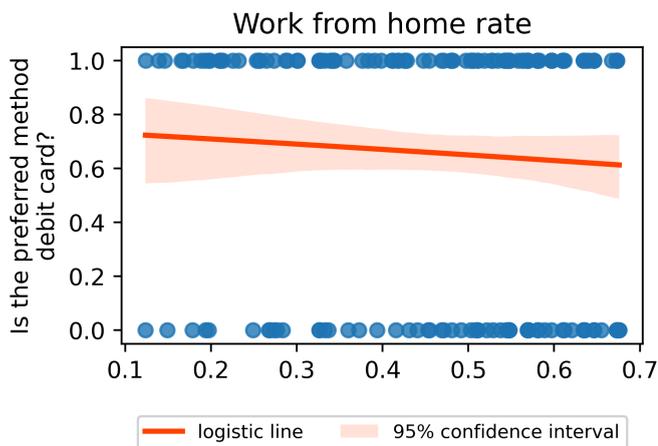

Odds Ratio (OR): 0.638941
p-value: 0.3732

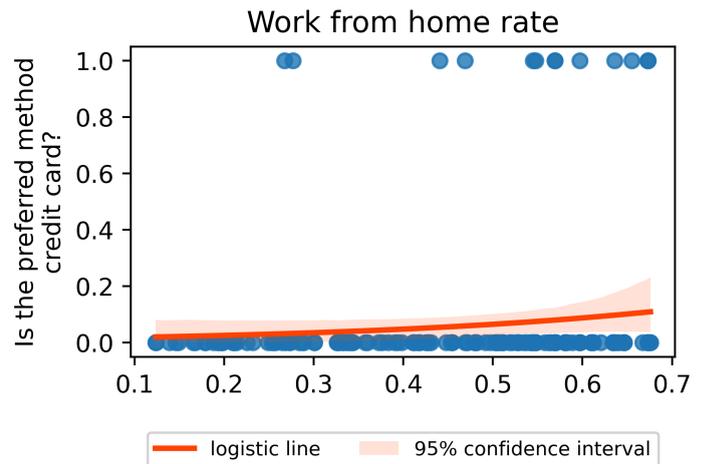

Odds Ratio (OR): 1.788727
p-value: 0.1685

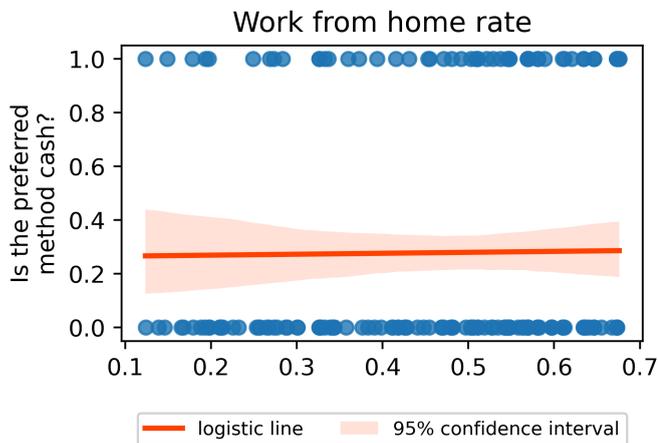

Odds Ratio (OR): 1.086667
p-value: 0.8679

*Acceptance $H_0$*: We reject the null hypothesis.

*Usage $H_0$*: We reject the null hypothesis.

*Preference $H_0$*: We fail to reject the null hypothesis.

The findings illustrate that merchants in locations with a larger number of people working from home are less likely to take cash payments and more likely to accept credit cards. This may be due to several factors related to changes in consumer behaviour and the evolving needs of merchants. With more people working remotely and conducting transactions online, there has been a shift towards electronic payments and a decline in the use of cash and merchants' acceptance reflects this shift. Furthermore, while the work-from-home rate does not appear to have any significant effect on the customers' usage of debit cards and contactless payments, it does have a positive correlation with statistical significance with the usage of credit cards. This might be because during the Covid-19 pandemic, when the census took place, white-collar workers were more likely to work-from-home than other working groups (Yeung, 2020). This consumer group might likely have higher disposable income, better credit scores, and greater access to credit cards. The ability to earn reward points incentivizes this consumer group to use the credit card where possible.

### 4.5 Relationship with population density data

*Motivation:*
Locations with greater population densities often have a larger demand for products and services and thus population density statistics may have a link with the payment methods that people use and shops accept in a certain area (Abbas, 2017). To meet the different requirements and preferences of their consumers, merchants in densely populated locations may be more inclined to accept a larger range of payment methods, such as credit and debit cards. Furthermore, because of the ease and speed they provide, people living in highly populated regions may be more likely to use card payments over cash.

*Acceptance $H_0$*: The population density rate in the given area has no effect on the vendor's acceptance of card payments or acceptance of cash.

*Usage $H_0$* : The population density rate in the given area has no effect on whether the most used payment method is contactless, debit card or credit card.

*Preference $H_0$* : The population density rate in the given area has no effect on whether the vendor prefers debit card payment, credit card payment or cash payment.

**Models:**

<u>Acceptance</u>

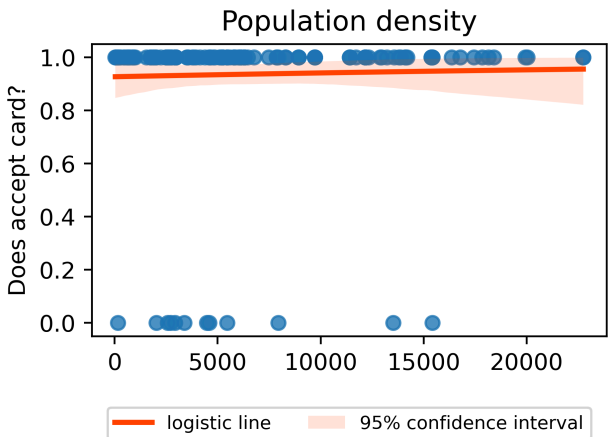

Odds Ratio (OR): 1.000023
p-value: 0.7082

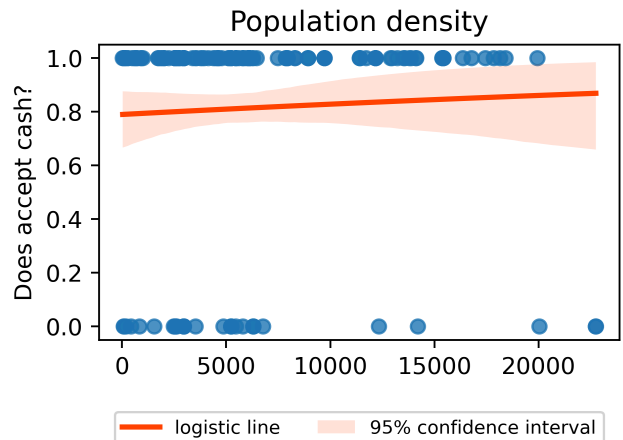

Odds Ratio (OR): 1.000025
p-value: 0.5217

## Usage

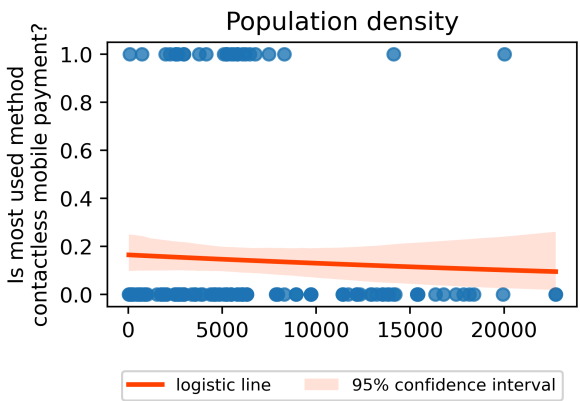

Odds Ratio (OR): 0.999972
p-value: 0.5268

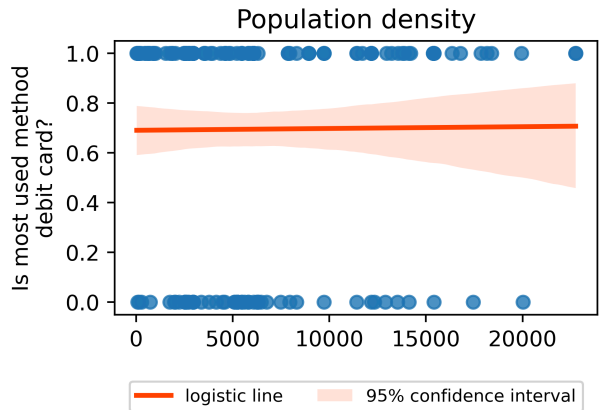

Odds Ratio (OR): 1.000003
p-value: 0.9126

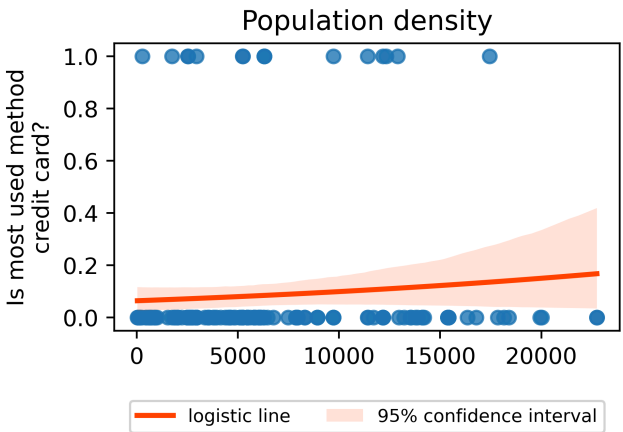

Odds Ratio (OR): 1.000047
p-value: 0.3087

## Preference

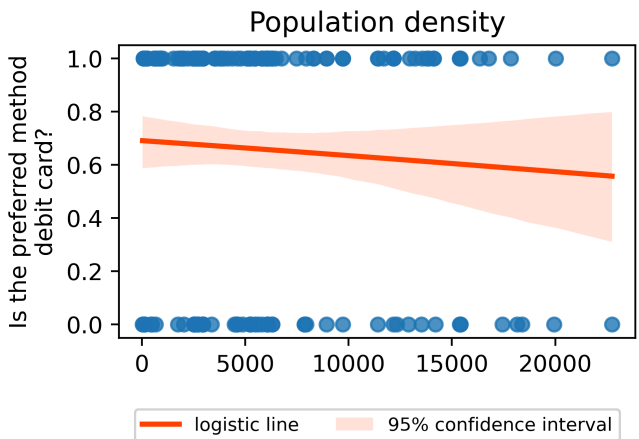

Odds Ratio (OR): 1.000054
p-value: 0.3971

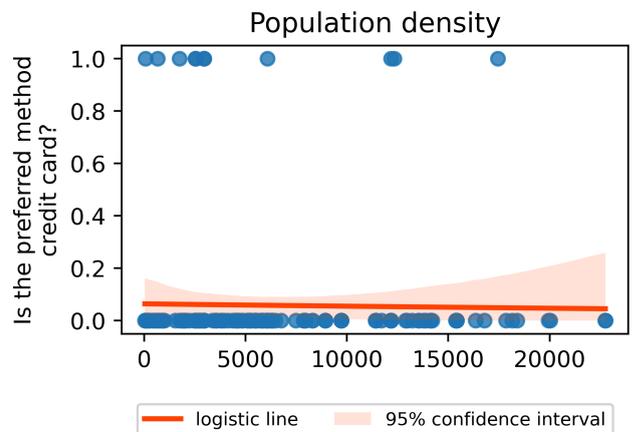

Odds Ratio (OR): 0.999984
p-value: 0.8013

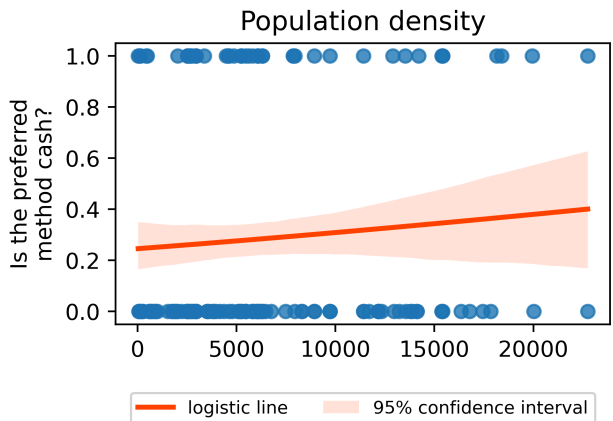

Odds Ratio (OR): 0.999919
p-value: 0.3071

*Acceptance $H_0$*: We fail to reject the null hypothesis.

*Usage $H_0$* : We fail to reject the null hypothesis.

*Preference $H_0$* : We fail to reject the null hypothesis.

*Comments*

The analysis failed to reject all the hypotheses, possibly due to no significant relationship between population density and acceptance and preference of payment methods. However, the population density of the output areas is not the best variable to use as the census boundaries do not necessarily correspond to natural or administrative boundaries. Having the output areas of similar area sizes could yield a different analysis outcome by providing a more accurate population density representation that closely reflects the true distribution of people within the study area.

### *5. Conclusion & Future Work*

We analysed the increase of card payments since 2012 and scrutinized significant barriers to card acceptance, focusing on the subtleties of certain implemented fees. We presented a case study of the regulation of interchange fees in the UK and EU and highlighted the merchant-specific outcomes. Our analysis of scheme fees implemented by Merchant Service Providers illustrated the current landscape of the well-established firms and the new start-up competitors that take a different approach in allowing merchants to take card payments. Finally, we observed relationships between census data in the given geography and the vendor's preferences. In particular, we observed interesting socio-demographic features suggesting that a younger population is associated with a higher usage of contactless payments; the higher the unemployment rate in the area, the more likely the vendor accepts cash; and the higher the work-from-home rate during the Covid-19 pandemic, the more likely it is that the most used payment method is a credit card.

One possible line of inquiry regarding future research is to create a thorough categorization model by utilising the different drivers that were examined in this study. While we looked into surface-level explanations for census and payment data, a single model that takes into account all factors has not yet been researched. There are obstacles in this approach, particularly when it comes to measuring societal drivers like preferences for card payments.. Deep learning methods and neural networks may be crucial in handling these attributes' complexity. Extending the study's scope to cover a wider range of payment methods and industries, as well as the retail sector and the cash-card payment combination, represents another avenue for future research. Further studies may also utilize an approximated expense and quantify specific obstacles related to each mode of payment for vendors, which would help with decision-making and give concrete data points to regulators who are looking to expand financial inclusion. Another productive line of inquiry is how new payment systems and interventions affect customer preferences and vendor acceptance. This will help to clarify how payment technologies are changing and what practical effects they will have on vendors and policymakers. Yet another future research direction could involve investigating how consumer preferences evolve over time in response to the introduction of new payment systems and which factors contribute to the adoption of different payment methods. The findings of such research would hold significant practical implications for policymakers seeking to foster the adoption of new payment technologies, promote financial accessibility, and help vendors lower transaction costs.

## Acknowledgements

The authors would like to thank Professor Tomaso Aste and the Systemic Risk Centre of the London School of Economics for their continued support of this line of research.

## Figure attributions

*Attributions:*

Credit card icons created by juicy_fish - Flaticon (https://www.flaticon.com/free-icons/credit-card)
Bank icons created by Freepik - Flaticon (https://www.flaticon.com/free-icons/bank)
Shopper icons created by Freepik - Flaticon (https://www.flaticon.com/free-icons/shopper)
Shop icons created by Nikita Golubev - Flaticon (https://www.flaticon.com/free-icons/shop)
Skyscraper icons created by Freepik - Flaticon (https://www.flaticon.com/free-icons/skyscraper)
Process icons created by Eucalyp - Flaticon (https://www.flaticon.com/free-icons/process)
Card reader icons created by Freepik - Flaticon (https://www.flaticon.com/free-icons/card-reader)
Digital wallet icons created by Freepik - Flaticon (https://www.flaticon.com/free-icons/digital-wallet)

# Appendix

# A  |  Acquirer Fee Analysis Data

| Name | Domestic cards | Non-domestic cards | AMEX fees | Flat rate |
|---|---|---|---|---|
| Zettle | 0.0175 | 0.0175 | 0.0175 | 0 |
| SumUp | 0.0169 | 0.0169 | 0.0169 | 0 |
| Square | 0.0175 | 0.0175 | 0.0175 | 0 |
| MyPos | 0.011 | 0.0285 | 0.0245 | 0 |
| Dojo | 0.014 | 0.0235 | 0.019 | 0.05 |
| Revolut Reader | 0.008 | 0.026 | | 0.02 |
| Takepayments | 0.006,0.025 | 0.006,0.025 | 0.006,0.025 | 0 |
| Worldpay | 0.0075,0.0275 | 0.0075,0.0275 | 0.0075,0.0275 | 0 |

The equation to calculate the loss of income: real_income = α - (α * γ) - φ
Where α is the value of the transaction, γ is the fee (as percentage) and φ is flat rate in pennies

Sources:

https://www.zettle.com/gb/pricing
https://help.sumup.com/en-GB/articles/4oI3qHHji2I2S9dyvRfec3-pricing-fees

https://squareup.com/help/gb/en/article/5068-what-are-square-s-fees#:~:text=Square's%20Payment%20Processi
ng%20Fees,-With%20the%20Square&text=1.75%25%20for%20each%20contactless%2C%20chip,Square%20I
nvoices%20and%20Virtual%20Terminal

https://www.mypos.com/en/pricing-and-fees
https://www.revolut.com/business/revolut-reader/
https://dojo.tech/pricing/
https://help.sumup.com/en-US/articles/Ya8cqbwB7i1Wc0HxjDnis-accepted-cards
https://startups.co.uk/payment-processing/best-small-business-credit-card-machines-readers
https://www.mobiletransaction.org/payzone-or-worldpay/